\documentclass[usenatbib,fleqn,useAMS]{mnras}

\usepackage{graphicx}
\usepackage{amsmath}
\usepackage{mathrsfs}
\usepackage{txfonts}
\usepackage{braket}
\usepackage{cancel}

\renewcommand\pi\upi
\renewcommand\le\oldleq
\renewcommand\ge\oldgeq

\newcommand\pdm\upartial
\newcommand\dm{\mathrm d}

\newcommand\rme{\mathrm e}
\newcommand\dirac\deltaup

\title[]{Self-consistent potential--density pairs of
thick disks and flattened galaxies.}

\author[An \& Evans]
{J.~An$^1$ and N.~W.~Evans$^2$\thanks{E-mail:~jin@kasi.re.kr;~nwe@ast.cam.ac.uk}
\\$^1$Korea Astronomy and Space Science Institute,
776 Daedeok-daero, Yuseong-gu, Daejeon, Republic of Korea (South);
\\$^2$Institute of Astronomy, University of Cambridge, Madingley Road,
Cambridge CB3~0HA.}

\pagerange{\pageref{firstpage}--\pageref{lastpage}}
\volume{000}\pubyear{2019}
\date{version \today.}

\begin{document}
\label{firstpage}
\maketitle

\begin{abstract}
We analyze the Miyamoto--Nagai substitution, which was introduced over
forty years ago to build models of thick disks and flattened
elliptical galaxies. Through it, any spherical potential can be
converted to an axisymmetric potential via the replacement of
spherical polar $r^2$ with $R^2 + ( a + \!\sqrt{z^2+b^2} )^2$, where
($R,z$) are cylindrical coordinates and $a$ and $b$ are constants. We
show that if the spherical potential has everywhere positive density,
and satisfies some straightforward constraints, then the transformed
model also corresponds to positive density everywhere. This is in
sharp contradistinction to substitutions like $r^2 \rightarrow R^2 +
z^2/q^2$, which leads to simple potentials but can give negative
densities.

We use the Miyamoto--Nagai substitution to generate a number of new
flattened models with analytic potential--density pairs. These include
(i) a flattened model with an asymptotically flat rotation curve,
which (unlike Binney's logarithmic model) is always non-negative for a
wide-range of axis ratios, (ii) flattened generalizations of the
hypervirial models which include Satoh's disk as a limiting case and
(iii) a flattened analogue of the Navarro--Frenk--White halo which has
the cosmologically interesting density fall-off of
(distance)$^{-3}$. Finally, we discuss properties of the prolate and
triaxial generalizations of the Miyamoto-Nagai substitution.
\end{abstract}

\section{Introduction}

Galaxies are flattened -- sometimes highly flattened. The
gravitational potential theory of spherical models is a done deal,
thanks to the work of Isaac Newton and his contemporaries. General
algorithms exist for constructing the potential--density pairs of
flattened systems~\citep{BT}. Especially if the density is stratified
on similar concentric ellipsoids, the potential can be written as a
single quadrature over elliptic integrals~\citep{EFE}. This though is
often inconvenient, especially for simulation work in which swift
orbit integration is a desideratum.

One way to construct a flattened analogue of a spherical model with 
potential $\psi(r)$ is to make the substitution
\begin{equation}
r^2 \rightarrow R^2 + z^2/q^2,
\end{equation}
where $q$ is a constant axis ratio.  This can never give a finite mass
model, as it is no longer possible for the asymptotic behaviour of the
potential to consist of a pure monopole. Nonetheless, if the potential
is a power-law, this has yielded some useful axisymmetric models with
flattish rotation curves~\citep[e.g.,][]{Ka93,Ev94,BT}.  The downside
of the substitution is that there is no guarantee that the density is
everywhere positive. This has to be checked \emph{a posteriori} via the
Poisson equation. This is a serious pitfall -- for example, the
potential of the famous NFW model~\citep{NFW}, if transformed in this
manner, never yields an everywhere positive mass density.

Another way to construct flattened analogues of spherical models was
introduced by \citet{MN75} and subsequently exploited by
\citet{NM76}, \citet{Sa80} and \citet{EB14}. Here, we take the
spherical potential $\psi(r)$ and make the substitution
\begin{equation}
r^2 \rightarrow R^2 + ( a + \!\sqrt{z^2+b^2} )^2,
\end{equation}
where $a$ and $b$ are constants. This has the advantage that a
spherical model with mass $M$ is transmogrified into a flattened model
with the same mass.  Despite providing a number of important Galactic
models~\citep[e.g.,][]{BT}, the transformation has not been
investigated systematically before. Here, we correct this lacuna,
prove some basic properties about the transformed models, and then
build some new models of cosmological interest, including flattened
haloes with density profiles falling like distance$^{-2}$ or
distance$^{-3}$. Finally, we show how the substitution can be modified
to build prolate or even triaxial models.

%\begin{figure*}
%\centering\includegraphics[width=.9\hsize]{mnd.eps}
%\caption{\label{fig:mnd}
%Some examples for the contour plots of the density profiles of
%the Miyamoto--Nagai disks. The levels are chosen logarithmically
%with respect to the central value, i.e.\ the reference values on the
%sidebar represent $\log_{10}[\rho^{\rm MN}(R,z)/\rho^{\rm MN}(0,0)]$.
%The same models are shown vertically on each column with increasingly
%larger scales whereas the contour plots on the same row are models
%with different parameters denoted by the label on the top.}
%\end{figure*}

\section{The Miyamoto--Nagai substitution}

Suppose that $(R,\phi,z)$ are cylindrical polar coordinates and
$\Phi(R,z)$ is the gravitational potential due to the density profile
$\rho(R,z)$, that is, $\nabla^2\Phi(R,z)=4\pi G\rho(R,z)$.  If we
substitute $z$ in $\Phi(R,z)$ by $h=a+\!\sqrt{z^2+b^2}$ \citep{MN75,NM76}
where $b\ge0$ and $a$ are some scaled constants, the resulting
function $\Phi^{\rm MN}(R,z)=\Phi(R,h)$ is still axisymmetric and also
reflection-symmetric about $z=0$. The density profile generating the
potential of $\Phi^{\rm MN}$ is then found from
\begin{align}\label{eq:dmn0}
\nabla^2\Phi^{\rm MN}
&=\frac1R\frac\pdm{\pdm R}\left(R\frac{\pdm\Phi^{\rm MN}}{\pdm R}\right)
+\frac{\pdm^2\Phi^{\rm MN}}{\pdm z^2}
\nonumber\\&=\frac1R\frac\pdm{\pdm R}\left(R\frac{\pdm\Phi}{\pdm R}\right)
+\left(1-\frac{b^2}{z^2+b^2}\right)\frac{\pdm^2\Phi}{\pdm h^2}
+\frac{b^2}{(z^2+b^2)^{3/2}}\frac{\pdm\Phi}{\pdm h}
\nonumber\\&=4\pi G\rho(R,h)-\frac{b^2}{\!\sqrt{z^2+b^2}}
\frac\pdm{\pdm h}
\left(\frac1{h-a}\frac{\pdm\Phi(R,h)}{\pdm h}\right),
\end{align}
and so, if the initial potential--density pair satisfies the conditions;
\begin{equation}
\rho\ge0,\quad
\frac{\pdm\Phi}{\pdm z}\ge0,\quad
\frac{\pdm^2\Phi}{\pdm z^2}\le0\quad
\text{for }\ z\ge a+b,
\end{equation}
the density profile resulting in the potential
$\Phi^{\rm MN}(R,z)=\Phi(R,h)$ is everywhere non-negative.
Note that this sufficient condition simply means that the initial
density is non-negative, and the vertical component of the gravitational
force acts towards the midplane and decreases as moving away from the plane,
all of which are characteristics of the gravitational field of any
centrally concentrated mass profiles. That is to say, if we start with
any physically reasonable potential--density pair, $\Phi(R,z)$ and
$\rho(R,z)$, the function $\Phi^{\rm MN}=\Phi(R,h)$ resulting from the
substitution $h=a+\!\sqrt{z^2+b^2}$ results in the potential of some
non-negative density profile.

%\begin{figure*}
%\centering\includegraphics[width=.9\hsize]{ebd.eps}
%\caption{\label{fig:ebd}
%Same as Figure~\ref{fig:mnd} except for the Evans--Bowden disks.}
%\end{figure*}

\subsection{Thickening up razor-thin disks}

If $b=0$, the density profile in equation (\ref{eq:dmn0}) for $z\ne0$
is simply $\rho^{\rm MN}(R,z)=\rho(R,a+|z|)$;
i.e.\ in the upper half space, the initial
spherical profile is displaced down by $a$ in the $z$ direction and
the density profile in the lower half space is its reflection about
the midplane. However, since
\begin{equation}
\frac{\pdm\Phi^{\rm MN}}{\pdm z}
=\frac z{\!\sqrt{z^2+b^2}}\frac{\pdm\Phi(R,h)}{\pdm h}
\stackrel{b=0}\Longrightarrow
\mbox{sgn}(z)\frac{\pdm\Phi(R,h)}{\pdm h}\Biggr\rvert_{h=a+|z|},
\end{equation}
where $\mbox{sgn}(x)=x/|x|$ is the sign function, the potential
$\Phi(R,a+|z|)$ is not differentiable in the $z$ direction on the
$z=0$ midplane. Rather, on the midplane, the two one-sided partial
derivatives with respect to $z$ actually have the same absolute values
but the opposite signs.  In fact, the situation corresponds to a
mathematical model for an infinitesimally-thin disk. Together with 
Gauss' theorem, the resulting system is understood to be the
superposition $\rho^{\rm MN}(R,z)=\rho(R,a+|z|)+\sigma(R)\dirac(z)$
(where $\dirac(z)$ is the Dirac delta function) of the ``displaced
reflection,'' $\rho(R,a+|z|)$ of the initial density and the
razor-thin disk whose surface density $\sigma(R)$ is given by
\begin{equation}\label{eq:rtdp}
\sigma(R)=\frac1{2\pi G}\lim_{z\to0^+}\frac{\pdm\Phi^{\rm MN}}{\pdm z}
=\frac1{2\pi G}\frac{\pdm\Phi(R,z)}{\pdm z}\Biggr\rvert_{z=a}.
\end{equation}
If $\Phi^\mathrm h(R,z)$ is an axisymmetric harmonic function,
the spatial function $\Phi^{\rm MN}(R,z)=\Phi^\mathrm h(R,a+|z|)$
is thus equivalent to the potential due to a pure razor-thin disk of
the surface density of equation (\ref{eq:rtdp}) with $\Phi=\Phi^\mathrm h$.
Then the potential function $\Phi^{\rm MN}(R,z)=\Phi^\mathrm h(R,h)$
following the full substitution $h=a+\!\sqrt{z^2+b^2}$ may be
seen as thickened counterparts of this infinitesimally-thin
(i.e.\ the $b=0$ case) disk.

\subsubsection{The Miyamoto--Nagai disk}

For example, if we choose $\Phi^\mathrm h\propto-r^{-1}$ where $r^2=R^2+z^2$,
the potential--density pair following the substitution is given by
\begin{equation}\label{eq:mnd}\begin{split}
\rho^{\rm MN}&=\frac M{4\pi}
\frac{b^2}{z_b^2[R^2+(a+z_b)^2]^{3/2}}\left(\frac a{z_b}
+\frac{3(a+z_b)^2}{R^2+(a+z_b)^2}\right);\\
\Phi^{\rm MN}&=-\frac{GM}{\bigl[R^2+(a+\!\sqrt{z^2+b^2})^2\bigr]^{1/2}},
\end{split}\end{equation}
where $z_b^2=z^2+b^2$ and $M$ is the finite total mass.
%In Figure~\ref{fig:mnd}, some isodensity contours for
%equation (\ref{eq:mnd}) are shown.
This is the same model as \citet{MN75}
originally investigated (henceforth the Miyamoto--Nagai disk).
At the $b=0$ limit, the Miyamoto--Nagai disk reduces to a razor-thin disk,
usually known as the Kuzmin disk\footnote{In the West,
the model had also been referred to as the model-$1$ of
the \citet{To63} disk family.}
\citep[after][]{Ku53,Ku56}; namely,
\begin{equation}
\sigma=\frac{aM}{2\pi}\frac1{(R^2+a^2)^{3/2}};\quad
\Phi^{\rm MN}=-\frac{GM}{\!\sqrt{R^2+(a+|z|)^2}}.
\end{equation}
On the other hand, the Miyamoto--Nagai disk at the $a=0$ limit
becomes a spherical model:
\begin{equation}\label{eq:plummer}
\rho^{\rm MN}=\frac{3b^2M}{4\pi}\frac1{(r^2+b^2)^{5/2}};\quad
\Phi^{\rm MN}=-\frac{GM}{\!\sqrt{r^2+b^2}},
\end{equation}
which is actually identified with the \citet{Pl11}
sphere.\footnote{The potential--density pair had been known earlier
as the regular solution to the index-5 Lane--Emden equation -- i.e.\
the $\gamma=\frac65$ polytrope \citep[e.g.,][]{Sc83}.}

\subsubsection{The Evans--Bowden disk}

If instead $\Phi^\mathrm h\propto\ln(r+z)$ (which is harmonic and
regular everywhere except on the $z\le0$ part of the $R=0$ axis),
the model we obtain after the Miyamoto--Nagai substitution is
(where $z_b^2=z^2+b^2$ again)
\begin{equation}\label{eq:ebd}\begin{split}
\rho^{\rm MN}&=\frac{\varv_\infty^2}{4\pi G}
\frac{b^2}{z_b^3\!\sqrt{R^2+(a+z_b)^2}}
\left(1+\frac{z_b(a+z_b)}{R^2+(a+z_b)^2}\right);
\\\Phi^{\rm MN}&=\varv_\infty^2
\ln\Biggl[\frac{a+z_b+\!\sqrt{R^2+(a+z_b)^2}}{2(a+b)}\Biggr],
\end{split}\end{equation}
with the zero point of the potential at the origin --
i.e.\ $\Phi^{\rm MN}(0,0)=0$.
%Some examples of the isodensity 
%contour for equation (\ref{eq:ebd}) are given in Figure~\ref{fig:ebd}.
Note the circular speed $\varv_\mathrm c$ due to the model is
\begin{equation}\label{eq:ebdcs}
\varv_\mathrm c^2(R)=R\frac{\pdm\Phi}{\pdm R}
=\varv_\infty^2\left(1-\frac{a+z_b}{\!\sqrt{R^2+(a+z_b)^2}}\right),
\end{equation}
that is, the midplane rotation curve is asymptotically flat.  This
model has been recently studied in detail by \citet{EB14}, who
introduced it as an extension of the Mestel disk and the isothermal
sphere.  This family is generally characterized by the density
fall-offs of $\rho^{\rm MN}\sim R^{-1}$ as $R\to\infty$ and $\sim
z^{-4}$ as $z\to\infty$ (for $b\ne0$), whilst the limiting razor-thin
disk (i.e.\ $b=0$) member of the model is found to be
\begin{equation}
\sigma=\frac{\varv_\infty^2}{2\pi G\!\sqrt{R^2+a^2}};\quad
\frac{\Phi^{\rm MN}}{\varv_\infty^2}
=\ln\Biggl[\frac{a+|z|+\!\sqrt{R^2+(a+|z|)^2}}{2a}\Biggr],
\end{equation}
which may be referred to as the (cored) \citet{Me63} disk\footnote{As
is noted by \citet{EdZ92}, this can be considered as the model-0
of the \citet{To63} disk family.} \citep[see also][]{Ly89}.
If additionally $a=b=0$, the surface density becomes simply
$\sigma\propto R^{-1}$ (i.e.\ the ``scale-free'' or singular
Mestel disk) and its midplane rotation curve is completely flat.

\section{Satoh's flattening substitution}

If the initial choice of potential--density pair is spherical such that
$\Phi(R,z)=\Phi^\mathrm s(r)$ where $r^2=R^2+z^2$, then the potential
function after the substitution $z\to h$ \citep{Sa80} is given by
$\Phi^{\rm MN}(R,z)=\Phi^\mathrm s(s)$ where
$s^2=R^2+h^2=r^2+a^2+b^2+2a\!\sqrt{z^2+b^2}$. Since the resulting
potential after the substitution is a function of the positions only
through $s$, all the equipotential surfaces also coincide with the
surfaces of constant $s$. At the same distance from the origin, the
value of $s$ is larger for the location on the $R=0$ axis than that on
the $z=0$ midplane. In other words, the substitution effectively
compresses the coordinate intervals in the $z$ direction relative to
the $R$ direction, and so the same change of $s$ is achieved by a
larger change of $R$ than $z$, indicating that the constant-$s$
surfaces are spaced more densely in $z$ than $R$. 

In addition, the gravity field,
$\bmath\varg=-\bmath\nabla\Phi^{\rm MN}
=-(\dm\Phi^\mathrm s/\dm s)\bmath\nabla s$ due to the potential
after the substitution is found to be
\begin{equation}\label{eq:mnf}
-\bmath\varg=\left(\frac Rs\hat{\bmath e}_R+
\frac{z\hat\Bbbk}{\!\sqrt{z^2+b^2}}\frac hs\right)
\frac{\dm\Phi^\mathrm s}{\dm s}
=\left(\bmath r+\frac{az\hat\Bbbk}{\!\sqrt{z^2+b^2}}\right)
\frac{GM^\mathrm s(s)}{s^3},
\end{equation}
where $\bmath r=R\hat{\bmath e}_R+z\Bbbk$ is the radius vector and
\begin{equation}
M^\mathrm s(r)=\frac{r^2}G\frac{\dm\Phi^\mathrm s(r)}{\dm r}
%=4\pi\!\int_0^r\!\dm r\,r^2\rho(r)=\frac{4\pi}3r^3\bar\rho(r)
\end{equation}
is the mass enclosed within the sphere of the radius $r$ for the
initial choice of the spherical model, whereas
$\hat{\bmath e}_R=\bmath\nabla R=\pdm\bmath r/\pdm R$
and $\hat\Bbbk=\bmath\nabla z=\pdm\bmath r/\pdm z$ are
the unit vectors in the increasing $R$ and
$z$ coordinate directions.  According to equation (\ref{eq:mnf}), if
$a>0$, the direction of the gravitational force is more inclined
vertically than pointing directly to the centre.  (If $a=0$, the force
points radially everywhere and the density is actually spherical; in
fact, the substitution then results in the simple softening
substitution, $r^2\to r^2+b^2$.) That is to say, everywhere, the slope
of the normal vectors to the equipotential surfaces (which is also the
surfaces of constant $s$) are steeper than that of the radius vector.
All these arguments together indicate that the equipotential surfaces
for the family of this potential after the substitution are oblate
(squashed vertically).

From the Poisson equation, the density profile resulting from the
substitution, $\rho^{\rm MN}=\nabla^2\Phi^{\rm MN}/(4\pi G)$ is shown
to be 
\begin{equation}\label{eq:dmns}\begin{split}
\rho^{\rm MN}(R,z)
&=\rho^\mathrm s(s)
+\frac{ab^2}{z_b^3}\frac{\bar\rho^\mathrm s(s)}3
+\frac{b^2h^2}{z_b^2}\frac{\bar\rho^\mathrm s(s)-\rho^\mathrm s(s)}{s^2}
\\&=\frac{R^2z_b^2+z^2h^2}{z_b^2s^2}\rho^\mathrm s(s)
+\frac{b^2}3\frac{aR^2+(a+3z_b)h^2}{z_b^3}\frac{\bar\rho^\mathrm s(s)}{s^2}
\end{split}\end{equation}
where $z_b^2=z^2+b^2$, $h=a+z_b$, and $s^2=R^2+h^2$,
with the spherical density functions defined to be
\begin{equation}\begin{split}
\rho^\mathrm s(r)&=\frac1{4\pi Gr^2}
\frac\dm{\dm r}\left(r^2\frac{\dm\Phi^\mathrm s(r)}{\dm r}\right)
=\frac1{4\pi r^2}\frac{\dm M^\mathrm s(r)}{\dm r},\\
\bar\rho^\mathrm s(r)&=\frac{3M^\mathrm s(r)}{4\pi r^3}
=\frac3{4\pi Gr}\frac{\dm\Phi^\mathrm s(r)}{\dm r}
=\frac3{r^3}\!\int_0^r\!\dm\tilde r\,\tilde r^2\rho(\tilde r);
\end{split}\end{equation}
that is, $\rho^\mathrm s(r)$ and $\bar\rho^\mathrm s(r)$ are the
density profile and the mean density within the sphere of the radius
$r$ corresponding to the initial choice of the spherical model.
Equation (\ref{eq:dmns}) also indicates that, if $\rho^\mathrm s\ge0$
(and integrable at the centre),
then the density profile due to $\Phi^\mathrm s(s)$ is non-negative.

Since $s=h=|a+b|\ge0$ at the origin,
\begin{equation}\label{eq:tms0}
\frac{\rho^{\rm MN}(R,z)}{\bar\rho^\mathrm s_0}\simeq1+\frac a{3b}
-\frac12\left(\frac{R^2}{R_0^2}+\frac{z^2}{z_0^2}\right)
\quad\text{(for $R^2,z^2\ll b^2$),}
\end{equation}
where $\bar\rho^\mathrm s_0=\bar\rho^\mathrm s(|a+b|)$,
$\rho^\mathrm s_0=\rho^\mathrm s(|a+b|)$, and
\begin{equation}\label{eq:tmsc}
\frac1{R_0^2}=\frac{a+5b}{b(a+b)^2}
\frac{\bar\rho^\mathrm s_0-\rho^\mathrm s_0}{\bar\rho^\mathrm s_0}
;\quad
\frac1{z_0^2}=\frac1{b^2}
\left(\frac{3a+5b}{a+b}\frac{\bar\rho^\mathrm s_0-\rho^\mathrm s_0}{\bar\rho^\mathrm s_0}+\frac ab\right).
\end{equation}
Here note, for any monotonic (i.e.\ radially non-increasing) spherical
density profile, we have $\rho^\mathrm s(\tilde r)\ge\rho^\mathrm
s(r)$ for $0\le\tilde r\le r$ and so $\bar\rho^\mathrm s(r)
=3r^{-3}\!\int_0^r\!\dm\tilde r\,\tilde r^2\rho^\mathrm s(\tilde r)
\ge3r^{-3}\!\int_0^r\!\dm\tilde r\,\tilde r^2\rho^\mathrm s(r)
=\rho^\mathrm s(r)$.  Therefore, the density in equation
(\ref{eq:dmns}) is regular at the centre for any \emph{physical}
starting spherical model unless $b=0$. The (squared) axis ratio of the
nearly spheroidal density contours is
\begin{equation}
\frac{R_0^2}{z_0^2}
=1+\frac ab\left(\frac{3a+7b}{a+5b}
+\frac{(a+b)^2}{b(a+5b)}
\frac{\bar\rho^\mathrm s_0}{\bar\rho^\mathrm s_0-\rho^\mathrm s_0}\right)
\ge1,
\end{equation}
and so the isodensity surfaces near the origin are oblate.
In general, the gradient of equation (\ref{eq:dmns}) is in the form of
\begin{equation}
\bmath\nabla\rho^{\rm MN}(R,z)
=-A[R,z;\rho(s)]\,\bmath r-\frac{aB[R,z;\rho(s)]\,z\hat\Bbbk}{\!\sqrt{z^2+b^2}},
\end{equation}
%
%\[A=\left(-\frac{k^2R^2+z^2h^2}{s^3}\rho'(s)
%+\frac{b^2}{s^3}\frac{aR^2+(a+5k)h^2}k\frac{\bar\rho(s)-\rho(s)}s\right)
%\frac1{k^2}\]\[
%B=\left[-\frac{k^2R^2+z^2h^2}{s^3}\rho'(s)+b^2\bar\rho(s)
%+\frac{b^2}{s^3}
%\frac{aR^2+(a+5k)h^2+2(a+k)s^2}k\frac{\bar\rho(s)-\rho(s)}s\right]
%\frac1{k^2}\]
%
where $A[R,z;\rho(s)]$ and $B[R,z;\rho(s)]$ are some functions of
positions determined by $\rho'(r)$, $\rho(r)$, and $\bar\rho(r)$
with $B\ge A\ge0$ provided that $\rho'(r)\le0$. Hence the density field
of equation (\ref{eq:dmns}) is monotonically decreasing everywhere
both in the $R$ and $z$ directions, and the isodensity surfaces are
even more squashed (i.e.\ the slopes of the normal vectors
are steeper) than the equipotential surfaces.

In the $b=0$ limit, the system is understood to be the superposition
$\rho^{\rm MN}(R,z)=\rho^\mathrm s(s)+\sigma(R)\dirac(z)$ of the
``displaced spherical density'' $\rho^\mathrm s(s)$ and the razor-thin
disk whose surface density is
\begin{equation}
\sigma(R)=\frac1{2\pi G}\lim_{z\to0^+}\frac{\pdm\Phi^{\rm MN}}{\pdm z}
=\frac a{2\pi GR_a}\frac{\dm\Phi^\mathrm s(r)}{\dm r}\Biggr\rvert_{r=R_a}
=\frac{aM(R_a)}{2\pi R_a^3},
\end{equation}
where $R_a^2=R^2+a^2$.
Although the razor-thin disk component is technically singular in
the three-dimensional density on the disk plane, the surface
density of the disk itself is still regular at the centre
with $\sigma(R)\simeq M^\mathrm s(a)/(2\pi a^2)
-[\bar\rho^\mathrm s(a)-\rho^\mathrm s(a)]R^2/a$ for $R^2\ll a^2$
as well as the central density of the displaced spherical component,
$\rho^\mathrm s(s)\simeq\rho^\mathrm s(a)+(\rho^\mathrm s)'(a)\,|z|$,
which is also finite.

%If $b=0$, the density profile in equation (\ref{eq:dmns}) for $z\ne0$
%is simply $\rho^{\rm MN}(R,z)=\rho^\mathrm s(s)$ where
%$s^2=R^2+(|z|+a)^2$; i.e.\ in the upper half space, the initial
%spherical profile is displaced down by $a$ in the $z$ direction and
%the density profile in the lower half space is its reflection with
%respect to the midplane. However, on the midplane, both $s$ and the
%potential $\Phi^\mathrm s(s)$ are not differentiable in the $z$
%direction due to the absolute value in $|z|$. By Gauss' Theorem, this
%implies that the resulting system is actually superposition
%$\rho^{\rm MN}(R,z)=\rho^\mathrm s(s)+\sigma(R)\dirac(z)$ (where
%$\dirac(z)$ is the Dirac delta) of a ``displaced spherical density''
%$\rho^\mathrm s(s)$ and a razor-thin disk whose surface density
%$\sigma(R)$ is given by
%
%\begin{equation}
%\sigma(R)=\frac1{2\pi G}\lim_{z\to0^+}\frac{\pdm\Phi^{\rm MN}}{\pdm z}
%=\frac a{2\pi Gs_0}\frac{\dm\Phi^\mathrm s(r)}{\dm r}\Biggr\rvert_{r=s_0}
%=\frac{aM(s_0)}{2\pi s_0^3},
%\end{equation}
%
%where $s_0=\!\sqrt{R^2+a^2}$.

\begin{figure*}
\centering\includegraphics[width=.9\hsize]{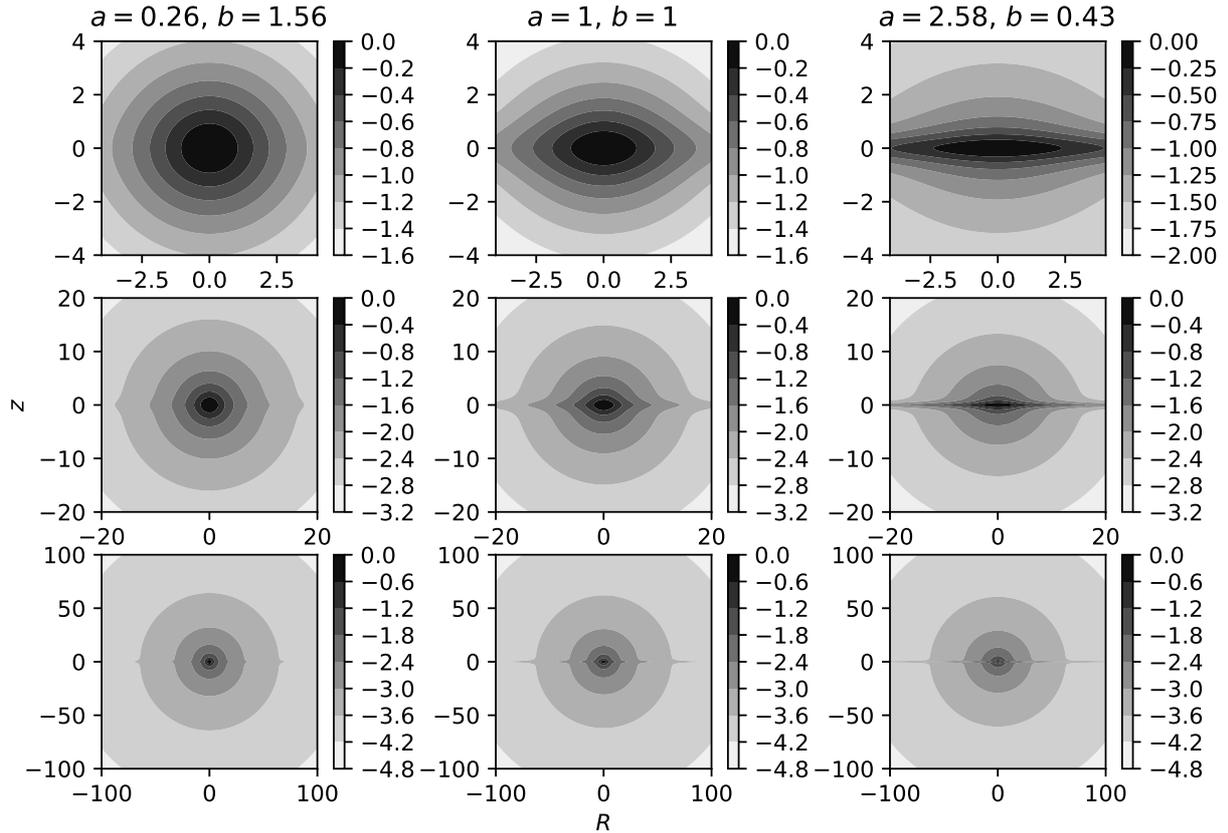}
\caption{\label{fig:fim}
%Same as Figure~\ref{fig:mnd} except for the flattened isothermal models.}
Some examples for the contour plots of the density profiles of
the flattened isothermal models. The levels are chosen logarithmically
with respect to the central value, i.e.\ the reference values on the
sidebar represent $\log_{10}[\rho^{\rm MN}(R,z)/\rho^{\rm MN}(0,0)]$.
The same models are shown vertically on each column with increasingly
larger scales whereas the contour plots on the same row are models
with different parameters denoted by the label on the top.}
\end{figure*}

\section{Examples}

\subsection{Power-Laws}
\label{sec:pl}

As the simplest example, first consider the scale-free spherical potential
given by
\begin{equation}\begin{split}
\Phi^\mathrm s(r)&=\Phi_0r_0^{1-\xi}
\begin{cases}\dfrac{r^{\xi-1}}{\xi-1}&(\xi\ne1)
\\\ln r&(\xi=1)\end{cases},\quad
\frac{\dm\Phi^\mathrm s(r)}{\dm r}=\frac{\Phi_0r_0^{1-\xi}}{r^{2-\xi}}\\
\rho^\mathrm s(r)&=\frac{\xi\Phi_0r_0^{1-\xi}}{4\pi Gr^{3-\xi}}
,\quad
\bar\rho^\mathrm s(r)
=\frac{3\Phi_0r_0^{1-\xi}}{4\pi Gr^{3-\xi}}.
\end{split}\end{equation}
Technically this model is ``physical'' only if $0\le\xi\le1$, but here
we do not introduce \emph{a priori} restrictions on the parameter $\xi$.
Rather, we examine the parameter range to yield a non-negative
density profile $\rho^{\rm MN}\ge0$ after the substitution
$r\to s=\!\sqrt{R^2+h^2}$ and $h=a+\!\sqrt{z^2+b^2}$.
From equation (\ref{eq:dmns}), the density profile $\rho^{\rm MN}(R,z)$
resulting in the potential of the form
$\Phi^{\rm MN}(R,z)=\Phi^\mathrm s(s)$ is given by
\begin{equation}\label{eq:gmn}
\frac{4\pi G\rho^{\rm MN}}{\Phi_0r_0^{1-\xi}}
=\frac{\xi}{s^{3-\xi}}+\frac{ab^2}{(z^2+b^2)^{3/2}s^{3-\xi}}
+\frac{(3-\xi)\,b^2h^2}{(z^2+b^2)s^{5-\xi}},
\end{equation}
whereas the right-hand side at the $b=0$ limit is replaced by
$[\xi+2a\dirac(z)]s^{\xi-3}$.
If $a\ge0$, this is non-negative for $0\le\xi\le3$, and
at the $\xi=3$ limit, we have a plane-parallelly stratified
density: namely,
\begin{equation}\begin{split}
\rho^{\rm MN}
&=\left[3+\frac{ab^2}{(z^2+b^2)^{3/2}}\right]\frac{\Phi_0}{4\pi Gr_0^2};
\\\frac{\Phi^{\rm MN}}{\Phi_0}&=\frac{R^2+(a+\!\sqrt{z^2+b^2})^2}{2r_0^2}.
\end{split}\end{equation}
If $a=0$, this is simply a harmonic potential of the constant density,
whilst the $b=0$ limit corresponds to the infinite uniform density plate
\citep[whose potential is a linear function of $z$ alone; see]
[sect.~22]{Ro92} superposed on top of it.
In any case, these limiting models are of only academic interests
and not at all realistic.

At the other end of the $\xi=0$ limit, the model is again recognized
as the Miyamoto--Nagai disk family (see eq.~\ref{eq:mnd}, which is
equivalent to eq.~(\ref{eq:gmn}) with $\xi=0$ and $\Phi_0r_0=GM$).
The Miyamoto--Nagai disk is also special among the family of models in
equation (\ref{eq:gmn}) in that it has a finite total mass
and it is the only member of the family whose density profile
falls off asymptotically with a different power-index in
the $R$ and $z$ direction (i.e.\
$\rho^{\rm MN}\sim R^{-3}$ and $\sim z^{-5}$ as $R,z\to\infty$,
whilst $\rho^{\rm MN}\sim s^{-(3-\xi)}\sim R^{-(3-\xi)}\sim z^{-(3-\xi)}$ for
all other members) and whose $b=0$ limit results in a pure
infinitesimally-thin disk.

\subsubsection{The flattened isothermal model}

Another notable case is for $\xi=1$, for which
\begin{equation}\begin{split}
\rho^{\rm MN}&=\frac{\varv_\infty^2}{4\pi Gs^2}
\left(1+\frac{ab^2}{(z^2+b^2)^{3/2}}
+\frac{2b^2(a+\!\sqrt{z^2+b^2})^2}{s^2(z^2+b^2)}\right);
\\\Phi^{\rm MN}&=\frac{\varv_\infty^2}2
\ln\left[\frac{R^2+(a+\!\sqrt{z^2+b^2})^2}{(a+b)^2}\right],
\end{split}\end{equation}
where $s^2=R^2+(a+\!\sqrt{z^2+b^2})^2$ with the scale constant
replace by $\Phi_0\to\varv_\infty^2$. Figure~\ref{fig:fim} shows
some examples for the isodensity contours of this case. As the asymptotic
fall-off of the density is like $\sim s^{-2}\sim R^{-2}\sim z^{-2}$,
this model may be considered as a ``flattened isothermal'' model. Note that
the circular speed behaves like
\begin{equation}
\varv_\mathrm c^2=R\frac{\pdm\Phi}{\pdm R}
=\frac{\varv_\infty^2R^2}{R^2+(a+z_b)^2}
\stackrel{z=0}{\Longrightarrow}\varv_\infty^2
\left(1-\frac{(a+b)^2}{R^2+(a+b)^2}\right),
\end{equation}
which actually is identical to that of the so-called logarithmic
potential of Binney \citep[sect.~2.3.2]{BT}.  In fact, the $a=0$ limit
of the current model is the same as the spherical limit for Binney's
potential \citep[or the softened isothermal sphere; eq.~2.12 of][]{Ev93}
\begin{equation}
\rho^{\rm MN}=\frac{\varv_\infty^2}{4\pi G(r^2+b^2)}
\left(1+\frac{2b^2}{r^2+b^2}\right);\quad
\Phi=\frac{\varv_\infty^2}2\ln\left(1+\frac{r^2}{b^2}\right).
\end{equation}
\emph{Unlike Binney's potential however, the isothermal sphere flattened
according to the Miyamoto--Nagai scheme is always non-negative and
better behaved for a much wider range of axis-ratios,} although the
equipotentials are not spheroidal any more.

\begin{figure}
\includegraphics[width=\columnwidth]{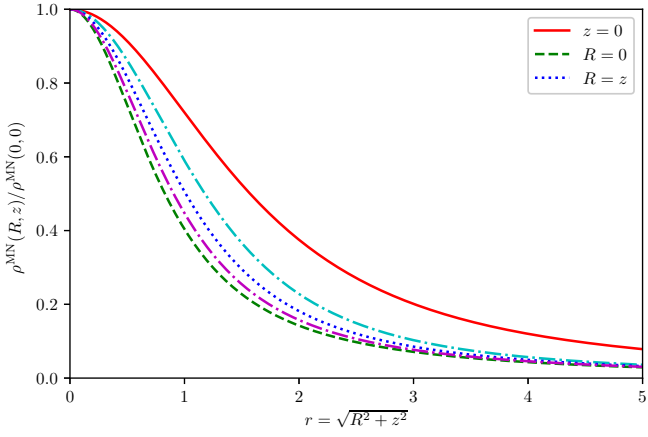}
\caption{\label{fig:fim_pr}
Density profiles along the one-dimensional radial ray
of the flattened isothermal model with $a=b=1$. The solid line is along
the major axis (on the plane) and the dashed line is along the minor axis
(on the symmetry axis). Also shown are the behaviours along some inclined
directions: the latitudes of 0\degr ($=$ along the major axis), 30\degr,
45\degr, 60\degr, and 90\degr ($=$ along the minor axis).
As the isodensity surfaces are ``oblate,'' the density decrease with
the increasing latitude at the same radius. Note that the density decays
noticeably slower on the plane than along the directions off plane.}
\end{figure}

Near the origin, we find that
\begin{multline}
\rho^{\rm MN}\simeq\frac{\varv_\infty^2}{4\pi Gb(a+b)^2}
\\\times\left[a+3b-\frac{a+5b}{(a+b)^2}R^2
-\frac{3a^2+9ab+10b^2}{2b^2(a+b)}z^2+\cdots\right],
\end{multline}
and so the axis ratio of the isodensity surface approaches to
\begin{equation}\label{eq:ar0}%\begin{split}
\frac{R_0}{z_0}
%&=\frac{(a+b)^{1/2}(3a^2+9ab+10b^2)^{1/2}}{\!\sqrt2b(a+5b)^{1/2}}\\&
=\!\sqrt{1+\frac{\chi(17+12\chi+3\chi^2)}{2(5+\chi)}},
%\end{split}
\end{equation}
as $R,z\to0$, where $\chi=a/b$. On the other hand, the asymptotic behaviours
of the density profile are found to be
\begin{equation}\begin{split}
\rho^{\rm MN}&\simeq\frac{\varv_\infty^2}{4\pi G}
\left(1+\frac{ab^2}{z_b^3}\right)\frac1{R^2}+\mathcal O(R^{-4})
\quad(R\to\infty),\\
\rho^{\rm MN}&\simeq\frac{\varv_\infty^2}{4\pi G}\frac1{z^2}
+\mathcal O(z^{-3})\quad(z\to\infty).
\end{split}\end{equation}
If $z=0$, then $z_b=b$ and so the axis ratio of the isodensity
surface asymptotically tends to $\to(1+a/b)^{1/2}\ge1$, which is closer
to the unity than the limiting value at equation (\ref{eq:ar0}) for all
$\chi>0$. However according to Figure \ref{fig:fim}, the shapes of the
isodensity surfaces at large radii are actually "Saturn-like"
(that is, a disk or ring on top of a sphere). Similarly the density
fall-offs of the model with $a=b=1$ shown in Figure \ref{fig:fim_pr}
also indicate the decay along the minor axis is noticeably slower than
those along the directions off plane (all of which tend to a common
behaviour implying approximate sphericity). This is understood from
the fact that, if $z/R\ne0$ is fixed, we have $z\to\infty$
and so $ab^2/z_b^3\to0$ as $R\to\infty$.

Lastly, the $b=0$ limit of the family is given by
\begin{equation}\begin{split}
\rho^{\rm MN}&=\frac{\varv_\infty^2}{4\pi G}\frac1{R^2+(a+|z|)^2}
+\frac{\varv_\infty^2}{2\pi G}\frac{a\dirac(z)}{R^2+a^2};
\\\Phi^{\rm MN}&=\frac{\varv_\infty^2}2
\ln\left(\frac{R^2+(a+|z|)^2}{a^2}\right).
\end{split}\end{equation}
The flat rotation curve here is actually due to the displaced singular
isothermal sphere part, and not to the razor-thin disk component,
which falls like $\sim R^{-2}$ as $R\to\infty$ and
may be considered as the model-$\frac12$ of the \citet{To63} disk.
We also note that the potential due to the razor-thin disk component
alone actually requires an elliptic integral (of the elliptic/spheroidal
coordinates) to write down \citep[Table~3, $m=1$]{EdZ92},
but the superposed potential of the current model only involves elementary
functions of the cylindrical coordinate components.

%\begin{figure*}
%\centering\includegraphics[width=.9\hsize]{sat.eps}
%\caption{\label{fig:sat}
%Same as Figure~\ref{fig:mnd} except for the Satoh disks.}
%\end{figure*}

\subsection{Hypervirial Models}

In order to obtain a model with a finite total mass, we need to start
with a spherical model with $\Phi^\mathrm s\sim r^{-1}$ as
$r\to\infty$. One such possibility is provided by the hypervirial
models of \citet{EA05} (also known as the \citet{Ve79} isochronous
spheres)
\begin{equation}\begin{split}
\Phi^\mathrm s(r)&=-\frac{GM}{(r^p+c^p)^{1/p}};\\
\bar\rho^\mathrm s(r)&
=\frac{3M}{4\pi}\frac1{r^{2-p}(r^p+c^p)^{1/p+1}};\\
\rho^\mathrm s(r)&=
\frac{M}{4\pi}\frac{(p+1)\,c^p}{r^{2-p}(r^p+c^p)^{1/p+2}}.
\end{split}\end{equation}
After the substitution, the density profile is then given by
\begin{multline}
\rho^{\rm MN}
%=\frac M{4\pi s^{2-p}(s^p+c^p)^{1/p+1}}
%\,\Biggl[\frac{ab^2}{(z^2+b^2)^{3/2}}
%+\frac{3b^2h^2}{z^2+b^2}\frac{s^{p-2}}{s^p+c^p}
%+\left(p+1+\frac{(2-p)b^2h^2}{(z^2+b^2)s^2}
%\right)\frac{c^p}{s^p+c^p}\Biggr]
=\frac M{4\pi s^{2-p}(s^p+c^p)^{1/p+1}}
\\\times\left[\frac{ab^2}{(z^2+b^2)^{3/2}}
+\frac{(p+1)\,c^p}{s^p+c^p}
+\frac{b^2h^2}{z^2+b^2}\frac{3s^p+(2-p)\,c^p}{s^2(s^p+c^p)}\right]
\end{multline}
which is nonnegative for $p\le2$. Here if $c=0$, the model
reproduces the Miyamoto--Nagai disk independent of $p$. If $c>0$,
the density falls off asymptotically like
$\rho^{\rm MN}\sim R^{-3}$ along the major axis ($z=0$),
whilst, along the minor axis ($R=0$), we find $\rho^{\rm MN}\sim z^{-(p+3)}$.

If $p=2$, we have some simplifications for
$s^p+c^p=s^2+c^2=r^2+a^2+b^2+c^2+2a\!\sqrt{z^2+b^2}=\mathcal S^2$ and so
\begin{multline}\label{eq:p2}
\rho^{\rm MN}=\frac M{4\pi\mathcal S^3}
\\\times\Biggl[\frac{ab^2}{(z^2+b^2)^{3/2}}
+\left(\frac{a^2b^2}{z^2+b^2}+\frac{2ab^2}{\!\sqrt{z^2+b^2}}
+b^2+c^2\right)\frac3{\mathcal S^2}\Biggr].
\end{multline}
Notably if $b^2+c^2\ge0$, this is still nonnegative everywhere.
In particular, if we choose $c^2=-b^2$, this model reduces to
\begin{equation}
\rho^{\rm MN}=\frac{ab^2M}{4\pi\mathcal S^3}
\left[\frac1{(z^2+b^2)^{3/2}}+\left(\frac a{z^2+b^2}
+\frac2{\!\sqrt{z^2+b^2}}\right)\frac3{\mathcal S^2}\right],
\end{equation}
where $\mathcal S^2=s^2+c^2=s^2-b^2=r^2+a^2+2a\!\sqrt{z^2+b^2}\ge0$.
%with the examples for the resulting isodensity contours shown
%in Figure~\ref{fig:sat}.
It turns out that this is equivalent to equation (8) of the \citet{Sa80},
who arrived at the same model via a somewhat roundabout
route.\footnote{Equation (8) of \citet{Sa80} is supposed to be
the $n\to\infty$ limit of
\begin{displaymath}
\rho_n=\frac{b^{2(n+1)}}{n!}\left(-\frac\pdm{\pdm(b^2)}\right)^n
\frac{\rho_\mathrm P}{b^2}\propto\frac{b^{2(n+1)}}{(r^2+b^2)^{n+5/2}},
\end{displaymath}
where $\rho_\mathrm P$ is that of the Plummer sphere (see
eq.~\ref{eq:plummer}), followed by the same flattening substitution.
However, the actual proper limit of the above expression is the Dirac
delta function (or equivalently the central point mass). Nevertheless
he had gotten a distinct model from the Miyamoto--Nagai disk thanks to
\begin{displaymath}
\Phi_\infty=-\frac{GM}r=-\frac{GM}{\!\sqrt{r^2+b^2}}
\left(1-\frac{b^2}{r^2+b^2}\right)^{-1/2}.
\end{displaymath}
If we replace $r^2+b^2$ in the right hand side of this last equation
by $s^2$, we also obtain the potential of the Satoh disk; viz.\
$\Phi=-GM/\!\sqrt{s^2-b^2}$.}
Also note that the minor-axis asymptotic density fall-off of the Satoh disk
is steeper like $\sim z^{-6}$ than all other cases ($b^2+c^2>0$) of
equation (\ref{eq:p2}), for which $\rho^{\rm MN}\sim z^{-5}$.

\begin{figure*}
\centering\includegraphics[width=.9\hsize]{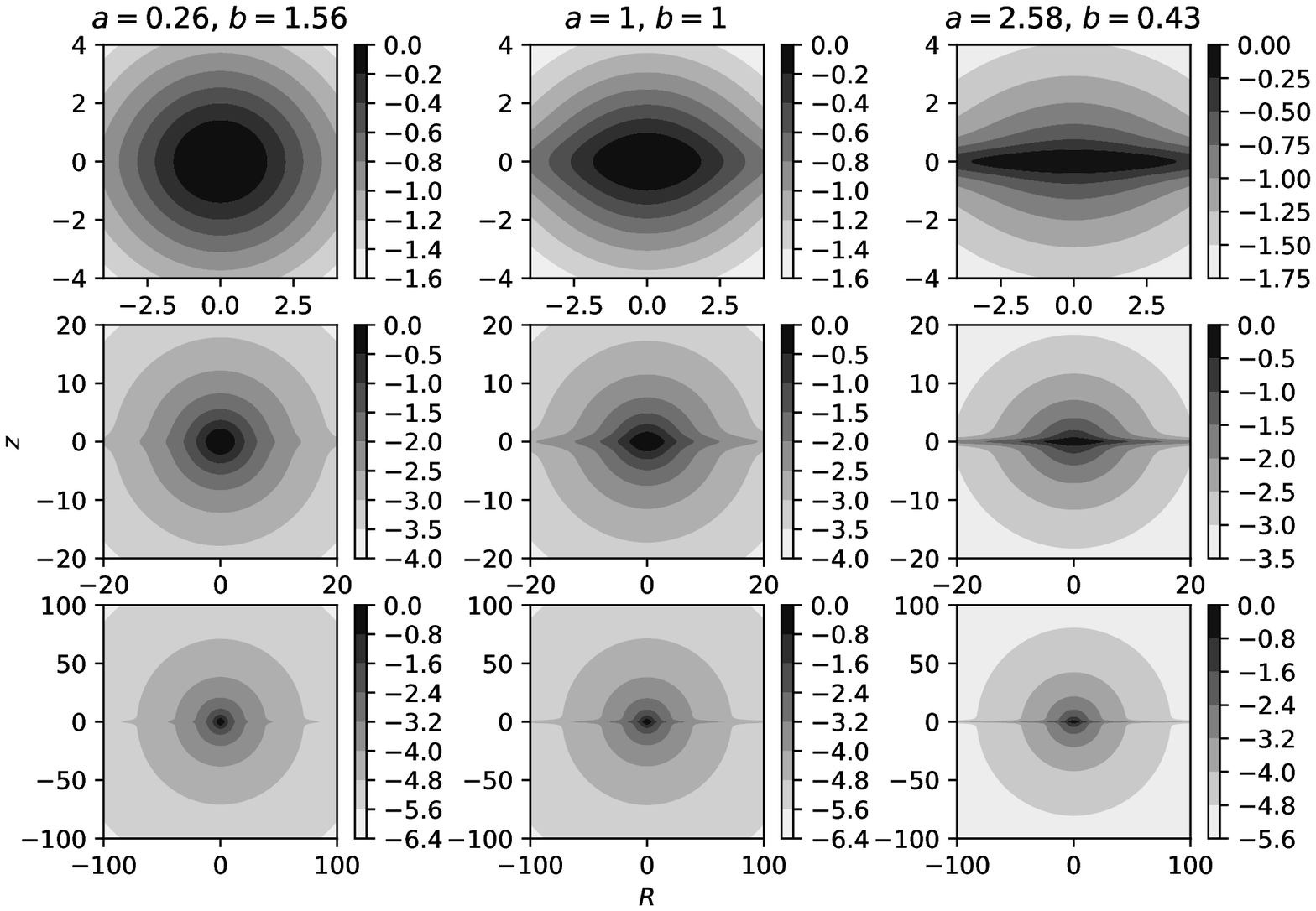}
\caption{\label{fig:xi3}
Same as Figure~\ref{fig:fim} except for the density profiles
in equation (\ref{eq:xi3}) with $r_0=(a+b)/\rme^{1/3}$.}
\end{figure*}

\section{Cosmological Halo Models}

So far, we have exploited the Miyamoto--Nagai substitution to provide
models with asymptotic density fall-off like $R^{-2}$ or $z^{-2}$ (the
flattened isothermal model), or like $R^{-3}$ along the major axis and
$z^{-5}$ or steeper along the minor axis (the flattened hypervirial
models including Satoh's model). Cosmological simulations suggest that
dark haloes may have densities that decay like distance$^{-3}$ \citep{NFW}.
It would be interesting to find flattened models
with this property. To do this, we first study the asymptotic
behaviour of models under the Miyamoto--Nagai substitution.

\subsection{Asymptotic Behaviour}

If the initial spherical model has finite total mass and the enclosed
mass approaches it like $M^\mathrm s(r)\simeq m_0-m_1(r/r_0)^{-\nu}$
($\nu>0$) as $r\to\infty$, we have asymptotic behaviour like
$\rho^\mathrm s(r)\simeq\nu m_1r_0^\nu/(4\pi r^{3+\nu})$ and
$\bar\rho^\mathrm s(r)\simeq 3m_0/(4\pi r^3)$. According to equation
(\ref{eq:dmns}), these indicate that the asymptotic density fall-offs
of the model after the Miyamoto--Nagai substitution are like
\begin{equation}\label{eq:rpasym}
\rho^{\rm MN}\simeq\frac{ab^2m_0}{4\pi z_b^3}\frac1{R^3}
;\qquad
\rho^{\rm MN}\simeq\frac{\nu m_1r_0^\nu}{4\pi}\frac1{z^{3+\nu}}
+\frac{3b^2m_0}{4\pi}\frac1{z^5}.
\end{equation}
That is to say, along the minor axis (i.e.\ the $R=0$ symmetry axis),
the resulting density profile falls off faster than $\sim z^{-3}$,
although the density on the $z=0$ plane always decays like $\sim R^{-3}$.
Instead, if we start with the model with the mass growing
like $M^\mathrm s(r)\simeq m_1(r/r_0)^\xi$ ($\xi>0$) as $r\to\infty$,
we have $\rho^\mathrm s(r)\simeq\xi m_1/(4\pi r_0^\xi r^{3-\xi})$
and $\rho^\mathrm s(r)\simeq 3m_1/(4\pi r_0^\xi r^{3-\xi})$, and so
the density profile after the substitution falls off with
the same power index as $\rho^\mathrm s$;
\begin{equation}\label{eq:rnasym}
\rho^{\rm MN}\simeq
\frac{m_1}{4\pi r_0^\xi}\left(\xi+\frac{ab^2}{z_b^3}\right)\frac1{R^{3-\xi}}
;\qquad
\rho^{\rm MN}\simeq
\frac{\xi m_1}{4\pi r_0^\xi}\frac1{z^{3-\xi}}.
\end{equation}

However, the cosmologically interesting asymptotic density fall-off of
$\rho^\mathrm s\sim r^{-3}$ \citep[e.g,][]{NFW} is actually the
borderline case. Unfortunately, if we start with the corresponding mass
model of $M^\mathrm s(r)\simeq m_1\ln(r/r_0)$, the resulting model
after the substitution exhibits asymptotic behaviour like
\begin{equation}\label{eq:r3asym}
\rho^{\rm MN}\simeq\frac{m_1}{4\pi R^3}
\left(1+\frac{ab^2}{z_b^3}\ln\frac R{r_0}\right)
;\qquad
\rho^{\rm MN}\simeq\frac{m_1}{4\pi z^3}.
\end{equation}
In other words, the asymptotic density fall-off on the plane is
strictly slower than that of $\sim R^{-3}$ due to the logarithmic
term (assuming $ab\ne0$). Nevertheless, we observe that the coefficient
of the $R^{-3}\ln R$ term in equation (\ref{eq:r3asym}) is linear in
$a$ whereas those of the $R^{-3}$ and $z^{-3}$ terms are independent of $a$.
Hence, if we consider the model given by the difference of
$\Phi^\mathrm s[\!\sqrt{R^2+(na+z_b)^2}]$ and
$n\Phi^\mathrm s[\!\sqrt{R^2+(a+z_b)^2}]$ where $z_b^2=z^2+b^2$
and $\Phi^\mathrm s(r)$ is the potential due to a spherical density profile
with the asymptotic behaviour like $\sim r^{-3}$, it is expected that
the resulting density profile behaves asymptotically like
both $\sim R^{-3}$ and $\sim z^{-3}$.

\begin{figure*}
\centering\includegraphics[width=.9\hsize]{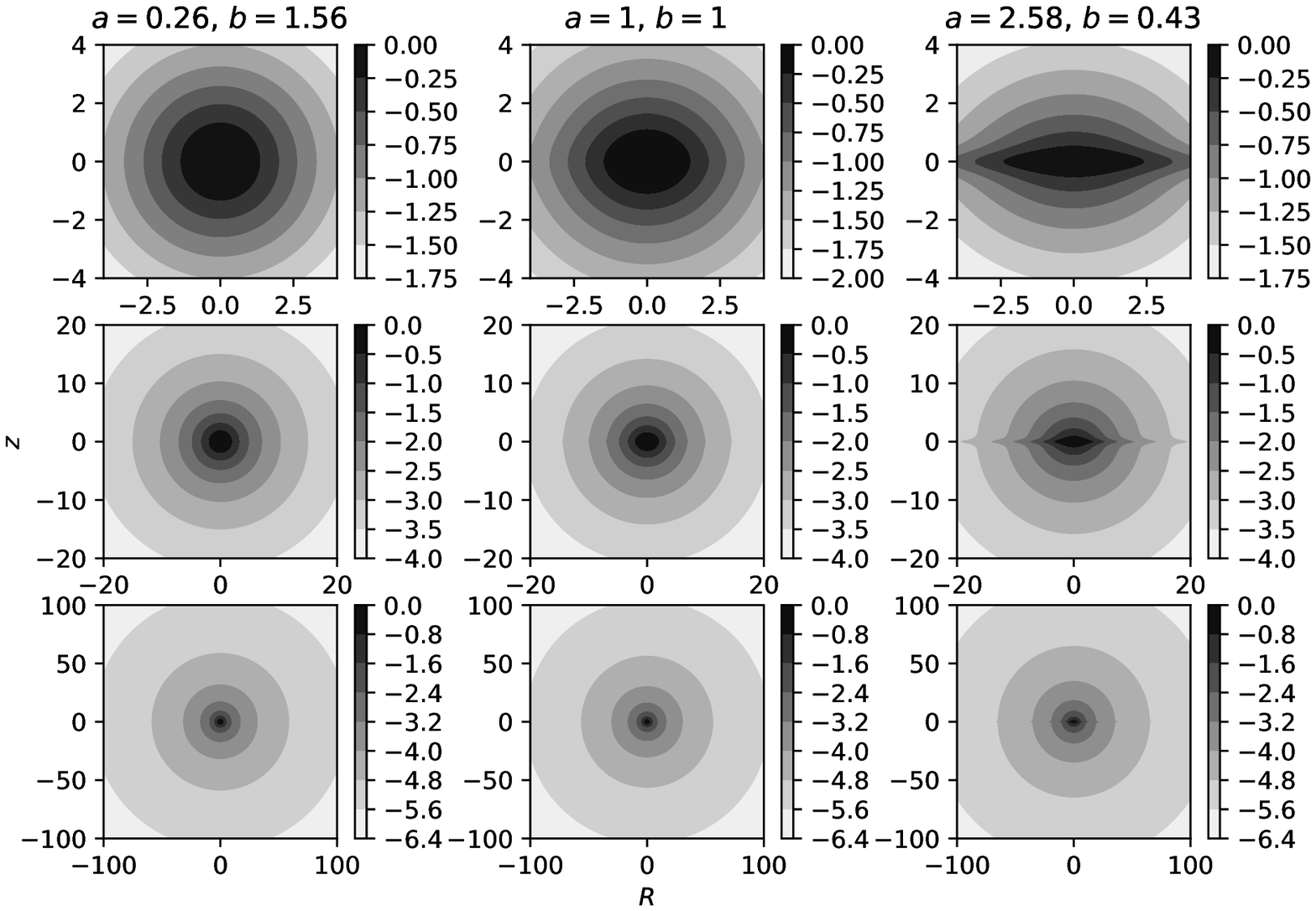}
\caption{\label{fig:dif}
Same as Figure~\ref{fig:fim} except for the density profiles
in equation (\ref{eq:dif}) with $r_0=(a+b)/\rme^{8/15}$.}
\end{figure*}

\subsection{The difference models with different $a$}

From equation (\ref{eq:dmns}) and $\dm\bar\rho^\mathrm s/\dm
r=3(\rho^\mathrm s-\bar\rho^\mathrm s)/r$, we find that
\begin{multline}\label{eq:pidx}
\frac\dm{\dm a}\!\left(\frac{\rho^{\rm MN}}a\right)
=\frac{R^2z_b^2+z^2h^2}{az_b^2s^2}\frac hs
\frac{\dm\rho^\mathrm s(s)}{\dm s}-\frac{\rho^\mathrm s(s)}{a^2}
\\-\frac{b^2h}{az_b^2}\left[1+\frac{(a-z_b)^2}{az_b}+\frac{5h^2}{s^2}\right]
\frac{\bar\rho^\mathrm s(s)-\rho^\mathrm s(s)}{s^2},
\end{multline}
which is negative provided that $\dm\rho^\mathrm s/\dm r\le0$ (and $a>0$).
That is to say, we have $\rho^{\rm MN}\approx Ca^\lambda$ locally
with $\lambda=\dm\ln\rho^{\rm MN}/\dm\ln a<1$, and so $\rho^{\rm MN}$ with
$a\to na$ is smaller than the $n$-times multiple of $\rho^{\rm MN}$
-- i.e.\ $C(na)^\lambda=n^\lambda\rho^{\rm MN}<n\rho^{\rm MN}$ -- for $n>1$.
It follows that, if $\Phi^\mathrm s(r)$ is the potential due to
a positive radially non-increasing spherical density profile,
then the function (where $n>1$ and $z_b^2=z^2+b^2$)
\begin{equation}\label{eq:ptdf}
\Phi^\mathrm d_n(R,z)=\frac{n\Phi^\mathrm s[\!\sqrt{R^2+(a+z_b)^2}]
-\Phi^\mathrm s[\!\sqrt{R^2+(na+z_b)^2}]}{n-1}
\end{equation}
is that of an axisymmetric non-negative density profile, $\rho^\mathrm
d_n(R,z)$, which may be found directly via the Poisson equation or
utilizing equation (\ref{eq:dmns}) with appropriate substitutions.
Furthermore, if the asymptotic behaviour of the original spherical
density profile is power-law-like, as in $\rho^\mathrm
s/\rho_0\simeq(r/r_0)^{-\nu}$ with $\nu<5$, then the asymptotic
behaviour of the resulting density profile also follows that like
$\rho^\mathrm d_n/\rho_0\simeq(R/r_0)^{-\nu}$ and
$\simeq(z/r_0)^{-\nu}$ along both major and minor axes (the isodensity
surfaces also becomes spherical in large radii). If $\nu>5$ on the
other hand, the asymptotic behaviour saturates at $\sim R^{-5}$ and
$\sim z^{-5}$ with
\begin{equation}\label{eq:rdasym}
\rho^\mathrm d_n\simeq
\frac{3b^2m_0}{4\pi R^5}\left(1+\frac{n(n+1)a^3}{2z_b^3}\right);\quad
\rho^\mathrm d_n\simeq\frac{3b^2m_0}{4\pi z^5},
\end{equation}
where $m_0$ is the total mass of the initial spherical model.  As for
the $\nu =5$ case, the terms like equation (\ref{eq:rdasym}) and
$\simeq\rho_0(r/R_0)^{-5}$ are of the same order and are simply added
together. For instance, if the difference model $\rho^\mathrm
d_n=n\rho^{\rm MN}(a)-\rho^{\rm MN}(na)$ is based on $\rho^{\rm MN}$
of equation (\ref{eq:p2}), the asymptotic behaviour of $\rho^\mathrm
d_n$ is found to be
\begin{equation}
\rho^\mathrm d_n\simeq
\frac{3M}{4\pi R^5}\left(b^2+c^2+\frac{n(n+1)a^3b^2}{2z_b^3}\right)
;\quad\rho^\mathrm d_n\simeq\frac{3(b^2+c^2)M}{4\pi z^5},
\end{equation}
whilst $\rho^\mathrm s\simeq3c^2M/(4\pi r^5)$.

\subsection{The differentiation model}

Given that equation (\ref{eq:pidx}) is always negative and the derivatives
are linear, if we consider the function given by
\begin{equation}\label{eq:dfm}\begin{split}
\Phi^\mathrm d_1(R,z)&=-a^2\frac\dm{\dm a}
\frac{\Phi^\mathrm s[\!\sqrt{R^2+(a+z_b)^2}]}a
\\&=\Phi^\mathrm s(s)-\frac{a(a+z_b)}{\!\sqrt{R^2+(a+z_b)^2}}
\frac{\dm\Phi^\mathrm s(s)}{\dm s},
\end{split}\end{equation}
where $\Phi^\mathrm s(r)$ is again a potential of a spherical density
profile, it becomes the potential due to the axisymmetric density
profile of $\rho^\mathrm d_1(R,z)=-a^2(\dm/\dm a)(\rho^{\rm MN}/a)$
where $\rho^{\rm MN}$ corresponds to the density profile for
$\Phi^\mathrm s[\!\sqrt{R^2+(a+z_b)^2}]$ (eq.~\ref{eq:dmns}). Thanks
to equation (\ref{eq:pidx}), $\rho^\mathrm d_1$ is non-negative if
$\Phi^\mathrm s(r)$ is a potential due to a positive radially
non-increasing density profile. In addition, the $a=0$ limit of the
model simply results in the same $a=0$ limit of $\rho^{\rm MN}$. In
fact, this model is the $n=1$ limit of the difference model discussed,
the fact of which can be proven by taking the limit of equation
(\ref{eq:ptdf}) as $n\to1$ by means of L'H\^opital's rule. Then the
asymptotic behaviour of the density profile is given by the same $n=1$
limit of the preceding discussion, namely, $\rho^\mathrm d_1\sim
R^{-\min(5,\nu)}$ and $\sim z^{-\min(5,\nu)}$ if we start with
$\rho^\mathrm s\sim r^{-\nu}$.

\begin{figure}
\centering\includegraphics[width=\columnwidth]{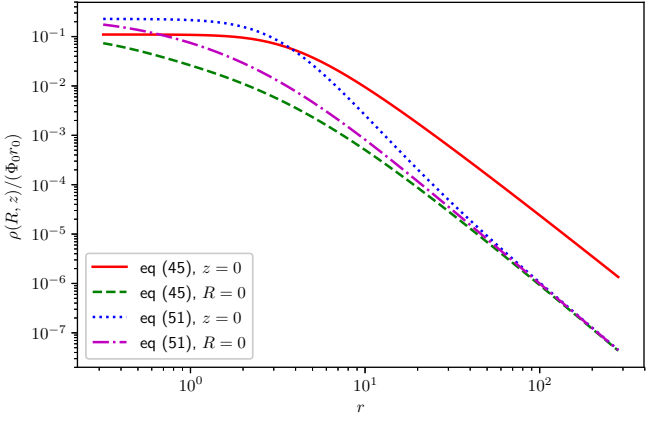}
\caption{\label{fig:xi3-dif}
Density profiles along the major and minor axes for the models
in eqs (\ref{eq:xi3}) and (\ref{eq:dif}). For both cases,
$a=2.58$ and $b=0.43$ but $r_0=\rme^{-1/3}(a+b)\approx2.157$ for
eq.~\ref{eq:xi3} and $r_0=\rme^{-1/3}(a+b)\approx1.766$ for eq.~\ref{eq:dif}.}
\end{figure}

\subsection{Models with the $\lowercase{\sim r^{-3}}$ density fall-offs}

If we starts with the potential of any spherical density profile that
falls off like $\sim r^{-3}$ as $r\to\infty$, the procedure outlined
so far will provided us with an analytic potential--density pair for a
``flattened'' axisymmetric system (becoming rounder at large radii)
with the asymptotic behaviour like $\sim R^{-3}$ and $\sim z^{-3}$.
Every calculation required to find the explicit expressions for the
potential--density pair is purely algebraic (including
differentiation) provided that the starting potential is also an
algebraic function.  That is to say, it is possible to
get a flattened density profile falling off like $\sim r^{-3}$ whose
potential is also expressible only using algebraic functions of the
cylindrical coordinates if we start with the potential for
density profiles of the form $\rho^\mathrm
s\propto1/[r^\alpha(c^p+r^p)^{(3-\alpha)/p}]$ with $p=1,2$ and
$\alpha=0,1,2$ \citep{AZ13} including the NFW profile
\citep[$\alpha=p=1$;][]{NFW} for which $\Phi^\mathrm
s(r)\propto-\ln(1+r/c)/r$.  However, the expressions do become rather
unwieldy, although their numerical implementations would be, if
tedious, trivial.

We prefer to begin with a starting potential as simple as
possible: consider the function
\begin{equation}\begin{split}
\Phi^\mathrm s(r)&=-\frac{\Phi_0r_0}r\left[\ln\left(\frac r{r_0}\right)+1\right]
;\quad\frac{\dm\Phi^\mathrm s}{\dm r}=
\frac{\Phi_0r_0}{r^2}\ln x
\\\rho^\mathrm s(r)&=\frac{\Phi_0r_0}{4\pi Gr^3};\quad
\bar\rho^\mathrm s(r)=\frac{3\Phi_0r_0}{4\pi Gr^3}\ln x
;\quad\frac{\dm\rho^\mathrm s(r)}{\dm r}=-\frac{3\Phi_0r_0}{4\pi Gr^4}
\end{split}\end{equation}
where $x=r/r_0$. This is not a physical potential since the
$r^{-3}$-cusp is not integrable (and also $\bar\rho<0$ for $x<1$ as
well as $\bar\rho<\rho$ for $x<\rme^{1/3}$). Nonetheless, this is
immaterial for our purpose because the Miyamoto--Nagai substitution
smears out the unphysical properties. Equation (\ref{eq:dmns}) is still
valid for the density profile after the Miyamoto--Nagai substitution,
and so we first arrive at the potential--density pair of the form
\begin{equation}\label{eq:xi3}\begin{split}
\Phi^{\rm MN}(R,z)&=-\frac{\Phi_0r_0}{s}
\left(1+\frac12\ln\frac{s^2}{r_0^2}\right)
\\\rho^{\rm MN}(R,z)&=\frac{\Phi_0r_0}{4\pi Gs^5}\Biggl(
R^2+\frac{z^2(a+z_b)^2}{z_b^2}
\\&\qquad
+b^2\frac{aR^2+(a+3z_b)(a+z_b)^2}{z_b^3}\ln\frac{s}{r_0}\Biggr)
\end{split}\end{equation}
where $z_b^2=z^2+b^2$ and $s^2=R^2+(a+z_b)^2$,
with the limiting cases;
\begin{equation}\begin{split}
\rho^{\rm MN}&\stackrel{a=0}=\frac{\Phi_0r_0}{4\pi G}
\frac{R^2+z^2+3b^2\ln(r_b/r_0)}{(R^2+z^2+b^2)^{5/2}}
\\\rho^{\rm MN}&\stackrel{b=0}\Rightarrow
\frac{\Phi_0r_0}{4\pi G}\left[\frac1{[R^2+(a+|z|)^2]^{3/2}}
+2\dirac(z)\frac{a\ln(R_a/r_0)}{(R^2+a^2)^{3/2}}\right]
\end{split}\end{equation}
where $R_a^2=R^2+a^2$ and $r_b^2=R^2+z^2+b^2$. In Figure~\ref{fig:xi3},
we have also shown some examples of the isodensity contours due to
the density profile in equation (\ref{eq:xi3}). This model may be considered
as an alternative $\xi=3$ limit of the power-law model discussed
in Sect.~\ref{sec:pl}. In addition we find that
$R^{-1}\pdm\rho^{\rm MN}(R,z)/\pdm R=-A-b^2B[3\ln(s/r_0)-1]$ and
$z^{-1}\pdm\rho^{\rm MN}(R,z)/\pdm z=-C-b^2D[3\ln(s/r_0)-1]$ where
$(A,B,C,D)$ are some non-negative algebraic functions of $(R,z,a,b)$
and so $r_0\le(a+b)\,\rme^{-1/3}\approx0.7165(a+b)$ is a sufficient
condition for the density profile to be non-negative and outwardly decreasing.
Near the origin, the density simply behaves like
\begin{equation}
\rho^{\rm MN}\simeq\frac{\Phi_0r_0}{4\pi Gb(a+b)^3}
\left[(a+3b)\ln x_0
-\frac12\left(\frac{R^2}{R_0^2}+\frac{z^2}{z_0^2}\right)+\cdots\right],
\end{equation}
where $x_0=(a+b)/r_0$ and
\begin{equation}\begin{split}
\frac1{R_0^2}&=\frac{(a+5b)(3\ln x_0-1)}{(a+b)^2};
\\\frac1{z_0^2}&=\frac1{b^2}\left[a+\frac{a^2+4ab+5b^2}{a+b}(3\ln x_0-1)\right],
\end{split}\end{equation}
whereas the asymptotic behaviours are like
\begin{equation}\begin{split}
\rho^{\rm MN}&\simeq\frac{\Phi_0r_0}{4\pi GR^3}
\left(1+\frac{ab^2\ln(R/r_0)}{(z^2+b^2)^{3/2}}\right)
+\mathcal O(R^{-5}\ln R)
&(R\to\infty);
\\\rho^{\rm MN}&\simeq\frac{\Phi_0r_0}{4\pi Gz^3}
\left(1-\frac{3a}z\right)
+\mathcal O(z^{-5}\ln z)
&(z\to\infty).
\end{split}\end{equation}

Finally, equation (\ref{eq:dfm}) based on this model then results in
\begin{equation}
\Phi^\mathrm d_1(R,z)=-\frac{\Phi_0r_0}s\left(1
+\frac12\frac{R^2+(2a+z_b)(a+z_b)}{R^2+(a+z_b)^2}
\ln\frac{s^2}{r_0^2}\right)
\end{equation}
and the corresponding density profile becomes
\begin{multline}\label{eq:dif}
\rho^\mathrm d_1(R,z)=\frac{\Phi_0r_0}{4\pi G}\Biggl\{\frac1{s^3}
+\frac{3a(a+z_b)}{s^5}\frac{R^2z_b^2+z^2(a+z_b)^2}{z_b^2s^2}
\\+\frac{ab^2(a+z_b)}{z_b^2s^5}
\left[1+\frac{(a-z_b)^2}{az_b}+\frac{5(a+z_b)^2}{s^2}\right]
\left(3\ln\frac s{r_0}-1\right)\Biggr\}
\end{multline}
with the $b=0$ limit being
\begin{multline}
\rho^\mathrm d_1(R,z)\stackrel{b=0}\Rightarrow
\frac{\Phi_0r_0}{4\pi G}\Biggl\{
\frac{R^2+(4a+|z|)(a+|z|)}{[R^2+(a+|z|)^2]^{5/2}}
\\+2\dirac(z)\frac{a^3[3\ln(R_a/r_0)-1]}{(R^2+a^2)^{5/2}}\Biggr\}.
\end{multline}
Figure~\ref{fig:dif} shows some example isodensity contours
resulting from equation (\ref{eq:dif}), whilst Figure~\ref{fig:xi3-dif}
illustrates the behaviours of the density along the major and minor axes
for a particular case. Since the model is
specifically constructed in such a way, the asymptotic behaviours
of the density are indeed given by
\begin{equation}
\rho^\mathrm d_1\simeq\frac{\Phi_0r_0}{4\pi GR^3}+\mathcal O(R^{-5}\ln R)
;\quad
\rho^\mathrm d_1\simeq\frac{\Phi_0r_0}{4\pi Gz^3}+\mathcal O(z^{-5}\ln z),
\end{equation}
whereas the Taylor--Maclaurin series is
\begin{multline}
\rho^\mathrm d_1\simeq\frac{\Phi_0r_0}{4\pi G(a+b)^3}
\Biggl[1+\frac{a^2+4ab+b^2}{b(a+b)}L_3
\\\hfill
-\frac{3(3a^3+5b^3)+(3a^3+15a^2b+25ab^2+5b^3)L_{15/8}}{10b^3(a+b)^2}z^2
\\-\left(3+\frac{a^2+6ab+b^2}{b(a+b)}L_{15/8}\right)\frac{R^2}{2(a+b)^2}
+\cdots\Biggr],
\end{multline}
where $L_3=3\ln[(a+b)/r_0]-1$ and $L_{15/8}=15\ln[(a+b)/r_0]-8$.
This last series also implies that the density profile might be
increasing near the origin if $(a+b)/r_0$ is chosen too small, but
the examinations of $\pdm\rho^\mathrm d_1/\pdm R$ and
$\pdm\rho^\mathrm d_1/\pdm z$ indicate that
$r_0\le(a+b)\rme^{-8/15}\approx0.5866(a+b)$ is actually sufficient
to guarantee the positive non-increasing density profile.

\begin{figure*}
\centering\includegraphics[width=.9\hsize]{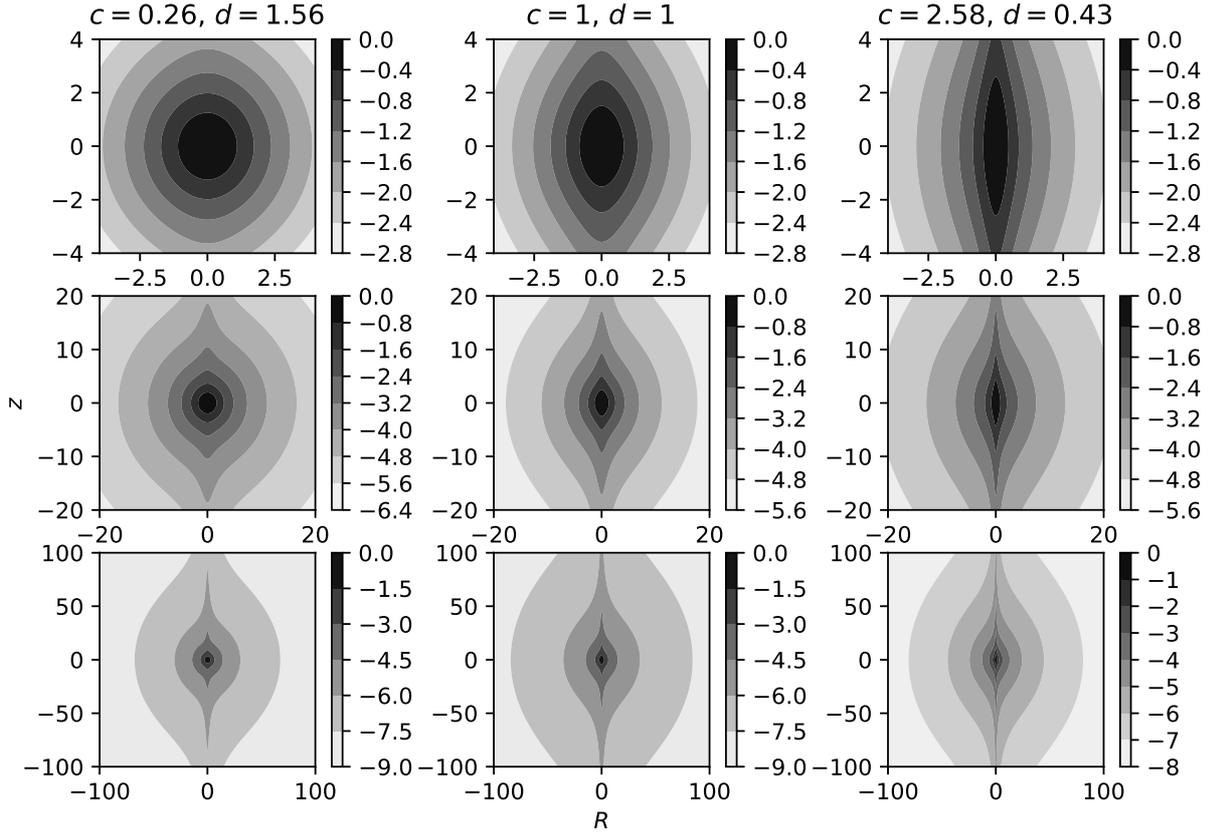}
\caption{\label{fig:mnsp}
Same as Figure~\ref{fig:fim} except for the density profiles
in equation (\ref{eq:mnsp}).}
\end{figure*}

\section{The Prolate and Triaxial Substitutions}

One possible generalization of the Miyamoto--Nagai substitution is
applying a similar transformation to the other coordinates.  For
instance, consider the transformation of a spherical function
$\Phi^\mathrm s(r)$ to $\Phi^{\rm gMN}(R,z)=\Phi^\mathrm s(t)$ where
$t^2=(c+R_d)^2+(a+z_b)^2$ and $(R_d^2,z_b^2)=(R^2+d^2,z^2+b^2)$.  With
$\Phi^{\rm gMN}(R,z)$ as the potential, the resulting gravity field is
given by $\bmath\varg=-\bmath\nabla\Phi^{\rm gMN} =-(\bmath\nabla
t)GM^\mathrm s(t)/t^2$ with
\begin{equation}
t(\bmath\nabla t)
=\left(1+\frac c{R_d}\right)R\hat{\bmath e}_R
+\left(1+\frac a{z_b}\right)z\Bbbk.
\end{equation}
The gravitational force points locally to a more vertically inclined
direction than that to the centre if $a/z_b>c/R_d$ and vice versa,
whilst it points radially towards the centre everywhere (i.e.\ the
spherical potential) if $a=c=0$.  The equipotential in the limit of
$R,z\to0$ is oblate if $a/b>c/d$ or prolate if $c/d>a/b$, whereas the
equipotentials as $R,z\to\infty$ become spherical. If $d=0$ and
$c\ne0$, we also find $\pdm t/\pdm R$ is non-zero finite and so the
gravitational force becomes undifferentiable along any path that
crosses the symmetry axis.  This behaviour is understood from the fact
that the corresponding density profiles are singular in this limit
(like $\sim R^{-1}$) on the $R=0$ axis.

The Poisson equation indicates that the axisymmetric density
profile generating the potential $\Phi^\mathrm s(t)$ can be expressed as
\begin{multline}\label{eq:gmn}
\rho^{\rm gMN}(R,z)
=\frac{\nabla^2\Phi^\mathrm s(t)}{4\pi G}=\rho^\mathrm s(t)
+\left[\frac{ab^2}{z_b^3}+\frac{c(R^2+2d^2)}{R_d^3}\right]
\frac{\bar\rho^\mathrm s(t)}3\\
+\left[\frac{b^2(a+z_b)^2}{z_b^2}+\frac{d^2(c+R_d)^2}{R_d^2}\right]
\frac{\bar\rho^\mathrm s(t)-\rho^\mathrm s(t)}{t^2},
\end{multline}
generalizing equation (\ref{eq:dmns}), which corresponds to the $c=d=0$ case.
It is easy to observe that this density profile is also always
non-negative if $\bar\rho^\mathrm s(r)\ge\rho^\mathrm s(r)\ge0$
(actually $\rho^\mathrm s(r)$ being non-negative and integrable at $r=0$ is
sufficient), and, from explicit calculation, we can furthermore demonstrate
that $\pdm\rho^{\rm gMN}/\pdm R\le0$ and $\pdm\rho^{\rm gMN}/\pdm z\le0$
if $(\rho^\mathrm s)'(r)\le0$
(and $\bar\rho^\mathrm s(r)\ge\rho^\mathrm s(r)\ge0$).

Since $t\ge t_0=\!\sqrt{(a+b)^2+(c+d)^2}\ge0$,
eqn (\ref{eq:gmn}) is also regular everywhere with
the Taylor-MacLaurin series is
\begin{equation}
\frac{\rho^{\rm gMN}(R,z)}{\bar\rho^\mathrm s(t_0)}\simeq
1+\frac{a}{3b}+\frac{2c}{3d}
-\frac12\left(\frac{R^2}{R_0^2}+\frac{z^2}{z_0^2}\right)+\cdots
\end{equation}
unless $bd=0$.\footnote{If $c=d=0$, see eqs.~(\ref{eq:tms0}-\ref{eq:tmsc}).
If $a=b=0$ on the other hand,
\begin{displaymath}
\frac{\rho^{\rm gMN}(R,z)}{\bar\rho^\mathrm s(t_0)}\simeq
1+\frac{2c}{3d}-\frac{(2c+5d)\Delta}{d(c+d)^2}\frac{z^2}2
-\left(\frac{4c}{3d}+\frac{(4c+5d)\Delta}{c+d}\right)\frac{R^2}{2d^2}+\cdots,
\end{displaymath}
where
$\Delta=[\bar\rho^\mathrm s(t_0)-\rho^\mathrm s(t_0)]/\bar\rho^\mathrm s(t_0)$
and $t_0=c+d$.}
Here,
\begin{equation}\begin{split}
\frac{b^2}{z_0^2}&=\frac ab+\frac{(a+b)(3ad+2bc+5bd)}{dt_0^2}
\frac{\bar\rho^\mathrm s(t_0)-\rho^\mathrm s(t_0)}{\bar\rho^\mathrm s(t_0)};\\
\frac{d^2}{R_0^2}&=\frac{4c}{3d}+\frac{(c+d)(ad+4bc+5bd)}{bt_0^2}
\frac{\bar\rho^\mathrm s(t_0)-\rho^\mathrm s(t_0)}{\bar\rho^\mathrm s(t_0)}.
\end{split}\end{equation}
If $d=0$ on the other hand, equation (\ref{eq:gmn}) reduces to
\begin{equation}
\rho^{\rm gMN}=\rho^\mathrm s(t)
+\left(\frac{ab^2}{z_b^3}+\frac cR\right)
\frac{\bar\rho^\mathrm s(t)}3
+\frac{b^2(a+z_b)^2}{z_b^2}
\frac{\bar\rho^\mathrm s(t)-\rho^\mathrm s(t)}{t^2},
\end{equation}
which exhibits an explicit $R^{-1}$-density singularity as $R\to0$
(i.e.\ the density along the symmetry axis diverges). This contrasts
to the $b=0$ case, for which a razor-thin disk develops.

\subsection{The Miyamoto--Nagai spindle}

The simplest potential--density pair that results from this
transformation is that due to the point mass potential $\Phi^\mathrm
s(r)=-GM/r$; viz.\
\begin{equation}
\Phi^{\rm gMN}=-\frac{GM}
{\bigl[(c+\!\sqrt{R^2+d^2})^2+(a+\!\sqrt{z^2+b^2})^2\bigr]^{1/2}};
\end{equation}
\begin{multline}\label{eq:gmn0}
\rho^{\rm gMN}=\frac{M}{4\pi t^3}
\Biggl[\frac{ab^2}{(z^2+b^2)^{3/2}}+\frac{c(R^2+2d^2)}{(R^2+d^2)^{3/2}}
\\+\frac3{t^2}\left(\frac{b^2(a+\!\sqrt{z^2+b^2})^2}{z^2+b^2}
+\frac{d^2(c+\!\sqrt{R^2+d^2})^2}{R^2+d^2}\right)\Biggr],
\end{multline}
where $t^2=(c+\!\sqrt{R^2+d^2})^2+(a+\!\sqrt{z^2+b^2})^2$. Obviously,
the $c=d=0$ case reduces to the Miyamoto--Nagai disk (or the Kuzmin disk
if $b=c=d=0$), whereas the $a=c=0$ case results in the Plummer sphere,
i.e.\ $\rho^{\rm gMN}\propto(r^2+b^2+d^2)^{-5/2}$. If $b=0$ and $a\ne0$,
the proper density model should additionally include the razor-thin disk
on the $z=0$ midplane with the surface density $\sigma(R)$ of
\begin{equation}
\sigma=\frac{aM}{2\pi\bigl[(c+\!\sqrt{R^2+d^2})^2+a^2\bigr]^{3/2}}.
\end{equation}
By contrast, if $d=0$ and $c\ne0$, eqn~(\ref{eq:gmn0}) contains
the $R^{-1}$ singular component -- $cM/(4\pi Rt^3)$ -- in it, and
with the simplest case that $a=b=d=0$ and $c\ne0$, only such a component
remains:
\begin{equation}
\Phi^{\rm gMN}=-\frac{GM}{\!\sqrt{(c+R)^2+z^2}}
;\quad\rho^{\rm gMN}=\frac{cM}{4\pi R[(c+R)^2+z^2]^{3/2}}.
\end{equation}

In Fig.~\ref{fig:mnsp}, we present the contour plots for some
examples of the pure spindle-like cases (viz.\ $a=b=0$) given by
\begin{equation}\label{eq:mnsp}\begin{split}
\Phi^{\rm gMN}&=-\frac{GM}
{\bigl[(c+\!\sqrt{R^2+d^2})^2+z^2\bigr]^{1/2}};
\\\rho^{\rm gMN}&=\frac{M}{4\pi t^3(R^2+d^2)}
\Biggl[\frac{c(R^2+2d^2)}{\!\sqrt{R^2+d^2}}
+\frac{3d^2(c+\!\sqrt{R^2+d^2})^2}{t^2}\Biggr].
\end{split}\end{equation}
We observe that the contours in the central region are quite similiar
to the $R$-$z$ flipped counterpart of the Miyamoto--Nagai disk, but
they fall off more slowly than those for the Miyamoto--Nagai disk.
The formal limit of the axis-ratio at the origin is actually found to
be $(z_0/R_0)^2=(1+\xi)(1+2\xi/3)$ where $\xi=c/d$ (which is close
enough to the reciprocal of the same ratio for the Miyamoto--Nagai disk,
$(R_0/z_0)^2=(1+\chi)(5+5\chi+\chi^2)/(5+\chi)$ where $\chi=a/b$
for a sufficiently small value of $\xi=\chi$).
In addition, the contours at large radii are more prominently
prolate bulge-like than the Miyamoto--Nagai disk case (resembling
a sphere plus a disk). This follows from the contrasting
asymptotic behaviour of the ``Miyamoto--Nagai spindle,'' for which
$\rho^{gMN}\sim z^{-3}$ and $\rho^{gMN}\sim R^{-4}$
(cf.\ $\rho^{MN}\sim R^{-5}$ and $\rho^{MN}\sim z^{-3}$
for the Miyamoto--Nagai disk).

\subsection{The Triaxial substitution}

Instead of restricting ourselves to the axisymmetric case, we may also
consider a triaxial generalization of the Miyamoto--Nagai
substitution; namely, $x_i\to a_i+(x_i^2+b_i^2)^{1/2}$ for all
$i=1,2,3$.  Whilst most of basic analyses of the models can proceed
similarly as before, here we only note the counterpart to
eqns~(\ref{eq:dmns}) and (\ref{eq:gmn}).  That is, the potential given
by $\Phi(x_1,x_2,x_3)=\Phi^\mathrm s(u)$ where
$u^2=\sum_{j=1}^3(a_j+x_{b,j})^2$ and $x_{b,j}^2=x_j^2+b_j^2$ can be
generated by the non-negative density profile of the form
\begin{equation}
\rho=\rho^\mathrm s(u)
+\sum_{j=1}^3\frac{a_jb_j^2}{x_{b,j}^3}\frac{\bar\rho^\mathrm s(u)}3
+\sum_{j=1}^3\frac{b_j^2(a_j+x_{b,j})^2}{x_{b,j}^2}
\frac{\bar\rho^\mathrm s(u)-\rho^\mathrm s(u)}{u^2},
\end{equation}
where $\rho^\mathrm s(r)$ and $\bar\rho^\mathrm s(r)$ are again
the local and average density profile corresponding to the spherical
potential $\Phi^\mathrm s(r)$.

\section{Conclusions}

This paper provides a systematic study of the \citet{MN75}
substitution. This is used in galactic dynamics to transform spherical
potential--density pairs to flattened ones.  Although introduced over
forty years ago to make the ubiquitous \cite{MN75} disk, the
substitution does not seems to have been thoroughly scrutinized
before, despite occasional model building \citep[e.g.,][]{Sa80,EB14}. 

The Miyamoto--Nagai substitution offers a number of advantages over the
much more familiar practice of transforming spherical equipotentials
to spheroidal or ellipsoidal ones.  Specifically, if the spherical
model has everywhere positive density, then the Miyamoto--Nagai
substitution is guaranteed to produce a physical model. Also, after
the Miyamoto--Nagai substitution, the potential retains the property
that it becomes spherical at large radii, meaning that the transformed
model can still have finite mass.

We have used the Miyamoto--Nagai substitution to provide some new
models. First, if applied to the isothermal sphere, it yields an
oblate isothermal model with an asymptotically flat rotation curve.
Prolate isothernmal models can be generated by the Miyamoto-Nagai
substitution applied to $R$ than than $z$.  In fact, the rotation
curve of these models is the same as for Binney's logarithmic model
\citep[e.g.,][]{Ev93,BT}, which is produced by the competing method of
converting spherical equipotentials to spheroidal ones with axis ratio
$q$. Binney's model ceases to generate physical densities once $q <
1/\sqrt{2} = 0.707$, and so cannot become very flattened.  However,
our prolate or oblate flattened isothermal model is always
non-negative and better behaved for a wider range of shapes.

Secondly, if we transform the hypervirial models (which includes the
Plummer sphere), we obtain a highly flattened family (which includes
the Satoh disk). The density along the major axis always falls
like $R^{-3}$, but along the minor axis it falls much more steeply,
between $z^{-3}$ and $z^{-5}$. Like the Satoh model itself, these are
useful for representing very highly flattened elliptical and lenticular
galaxies.

Third, we used the transformation to provide cosmological haloes,
inspired by the Navarro--Frenk--White profile \citep{NFW}. Here, we wish
to build flattened models with simple potential--density pairs that
have an asymptotic density fall-off like distance$^{-3}$. This proved
to be unexpectedly hard work, but can be done by taking the difference
between Miyamoto--Nagai transformed models (or equivalently,
differentiating with respect to the parameters in the
potential). Flattened or triaxial NFW-like models with analytic
potentials are hard to construct \citep[cf.][for a different
method]{BEB}, and we plan to return to this problem in a later
publication.

\section*{Acknowledgments}
NWE thanks the Korean Astronomy and Space Science Institute for their
hopitality during a working visit at which this paper was started.

\label{lastpage}
\end{document}